\documentclass[journal=nalefd,manuscript=article]{achemso}
\usepackage{chemformula} % Formula subscripts using \ch{}
\usepackage[T1]{fontenc} % Use modern font encodings
\usepackage{amsfonts}
\usepackage{bm}
\usepackage{multirow}
\usepackage{color}
\usepackage{multirow}
\usepackage{epsfig}
\usepackage{graphicx}
\usepackage[normalem]{ulem}
\usepackage{amsmath}
\usepackage{amssymb}
\usepackage{bm}
\usepackage{dcolumn}
\usepackage{braket}
\usepackage{bbold}
\usepackage{natbib}

\usepackage{array}
\usepackage{tabulary}
\usepackage{multirow}
\newcolumntype{C}[1]{>{\centering\arraybackslash}p{#1}}
\newcolumntype{L}[1]{>{\flushleft\arraybackslash}p{#1}}
\author{Botao Fu}
\affiliation{College of Physics and Electronic Engineering, Center for Computational Sciences, Sichuan Normal University, Chengdu, 610068, China}

\author{Run-Wu Zhang}
\affiliation
{Centre for Quantum Physics, Key Laboratory of Advanced Optoelectronic Quantum Architecture and Measurement (MOE), School of Physics, Beijing Institute of Technology, Beijing, 100081, China}
\alsoaffiliation
{Beijing Key Lab of Nanophotonics \& Ultrafine Optoelectronic Systems,School of Physics, Beijing Institute of Technology, Beijing, 100081, China}
\email{zhangrunwu@bit.edu.cn}

\author{Xiaotong Fan}
\affiliation
{Centre for Quantum Physics, Key Laboratory of Advanced Optoelectronic Quantum Architecture and Measurement (MOE), School of Physics, Beijing Institute of Technology, Beijing, 100081, China}
\alsoaffiliation
{Beijing Key Lab of Nanophotonics \& Ultrafine Optoelectronic Systems,School of Physics, Beijing Institute of Technology, Beijing, 100081, China}

\author{Si Li}
\affiliation{School of Physics, Northwest University, Xi'an 710127, China}
\alsoaffiliation{Shaanxi Key Laboratory for Theoretical Physics Frontiers, Xi'an 710127, China}

\author{Da-Shuai Ma}
\affiliation{Institute for Structure and Function $\&$ Department of Physics, Chongqing University, Chongqing 400044, China}

\author{Cheng-Cheng Liu}
\affiliation
{Centre for Quantum Physics, Key Laboratory of Advanced Optoelectronic Quantum Architecture and Measurement (MOE), School of Physics, Beijing Institute of Technology, Beijing, 100081, China}
\alsoaffiliation
{Beijing Key Lab of Nanophotonics \& Ultrafine Optoelectronic Systems,School of Physics, Beijing Institute of Technology, Beijing, 100081, China}
\email{ccliu@bit.edu.cn}

\title{2D Ladder polyborane: an ideal Dirac semimetal with a multi-field-tunable band gap}

\begin{document}
	%%%%%%%%%%%%%%%%%%%%%%%%%%%%%%%%%%%%%%%%%%%%%%%%%%%%%%%%%%%%%%%%%%%%%
	%% The "entry" environment can be used to create an entry for the

%	\begin{tocentry}
	
%	\includegraphics[width=8.2 cm,height=4.00 cm]{TOC}\label{For Table of Contents Only}

%	\end{tocentry}
		
\newpage

\begin{abstract}
Hydrogen, a simple and magic element, has attracted increasing attention for its effective incorporation within solids and powerful manipulation of electronic states. Here, we show that hydrogenation tackles common problems in two-dimensional borophene, e.g., stability and applicability. As a prominent example, a ladder-like boron hydride sheet, named as 2D ladder polyborane, achieves the desired outcome, enjoying the cleanest scenario with an anisotropic and tilted Dirac cone, that can be fully depicted by a minimal two-band tight-binding model. Introducing external fields, such as an electric field or a circularly-polarized light field can effectively induce distinctive massive Dirac fermions, whereupon four types of multi-field-driven topological domain walls hosting tunable chirality and valley indexes are further established. Moreover, the 2D ladder polyborane is thermodynamically stable at room temperature and supports highly switchable Dirac fermions, providing an ideal platform for realizing and exploring the various multi-field-tunable electronic states.


\end{abstract}

\newpage

The exploration of two-dimensional (2D) materials beyond graphene would offer great potential for future high-performance devices, sparking widespread interest in condensed matter physics and materials science. As graphene's counterparts for boron, a large variety of borophenes have been designed with abundant electronic structures involving insulators, semimetals, and metals\cite{PhysRevLett.112.085502,mannix2015synthesis,feng2016experimental,jacsxu2017two,zhang2020fully,bjPhysRevLett.118.096401}. Due to the shorter covalent radius and diversiform bonding nature of boron element, these materials are endowed with intriguing physical phenomena and potential applications\cite{wang2019review,xie2020two,ou2021emergence}, such as anisotropic metallicity\cite{lherbier2016electronic,PhysRevB.93.241405,mannix2018borophene}, phonon-mediated superconductivity\cite{PhysRevB.93.014502,PhysRevB.95.024505,Nanoscale2021LYan}, and high mechanical strength\cite{wang2016strain,zhang2017elasticity}.
Despite such merits, they confront a tough challenge that present bared boron sheets are prone to oxidation under ambient conditions, impeding their integration into practical applications.

Hydrogenation, as an ideal and feasible chemical passivation strategy, can effectively suppress the ambient oxidation of boron sheets\cite{xu2016hydrogenated,jiao2016two,wang2017new,wang2019band,xu2020realizing,kang2022substrate,pang2022first,nishino2017formation,hou2020ultrastable,li2021synthesis}. The recent experimental advances\cite{li2012elongation} in the boron analogues of polyenes offered a tremendous boost to the emerging field of partially-hydrogenated boron clusters. These boron cluster dihydrides (termed as B$_n$H$_2$-cluster) with various structures, having diverse bonding properties, play a key role in hydrogen-storage material fields. Moreover, broadening the choice of hydrogenated boron-based polymorphisms offers a potential arena for the eventual realization of device applications. However, the fascinating physical properties of such cluster dihydrides have been rarely studied.

On the other hand, through the self-assembly of small molecules or large clusters, several 2D materials have been successfully synthesized in experiments\cite{kumar2017molecular}. In particular, an intermixing phase of borophene is recently been prepared via the self-assembly of line defects\cite{liu2018intermixing}. Buoyed by the latest success of molecules self-assembly and aforementioned B$_n$H$_2$-clusters, in this work, we predict a 2D ladder-like hydroboron that is derived from the self-assembly of B$_n$H$_2$-clusters, named 2D ladder polyborane, which showcases as an ideal Dirac semimetal with an anisotropic and tilted Dirac fermion and highly switchable band gap.
First, we unveil that hydrogenation can substantially enhance the thermodynamic stability of ladder polyborane, and further drive a phase transition from normal semimetal into anisotropic and tilted Dirac semimetal by modifying the electron filling number.
Then, a minimal two-band tight-binding model is developed that is capable of reproducing whole features of Dirac bands, involving anisotropic Dirac point at the Fermi level and multiple saddle points near the Fermi level. Upon the combination of first principles calculation and low-energy effective $k{\cdot}p$ model, a vertical electric field and circularly polarized light field are simulated on ladder polyborane that can open the Dirac fermion with distinctive masses terms.
Finally, four types of multi-field-driven topological domain walls are illustrated which host topological interfacial states with tunable chirality and valley index.

%%%%%%%%%%%%%%%%%%
\section{RESULTS AND DISCUSSION}
\begin{figure}
\includegraphics[width=4.5 in]{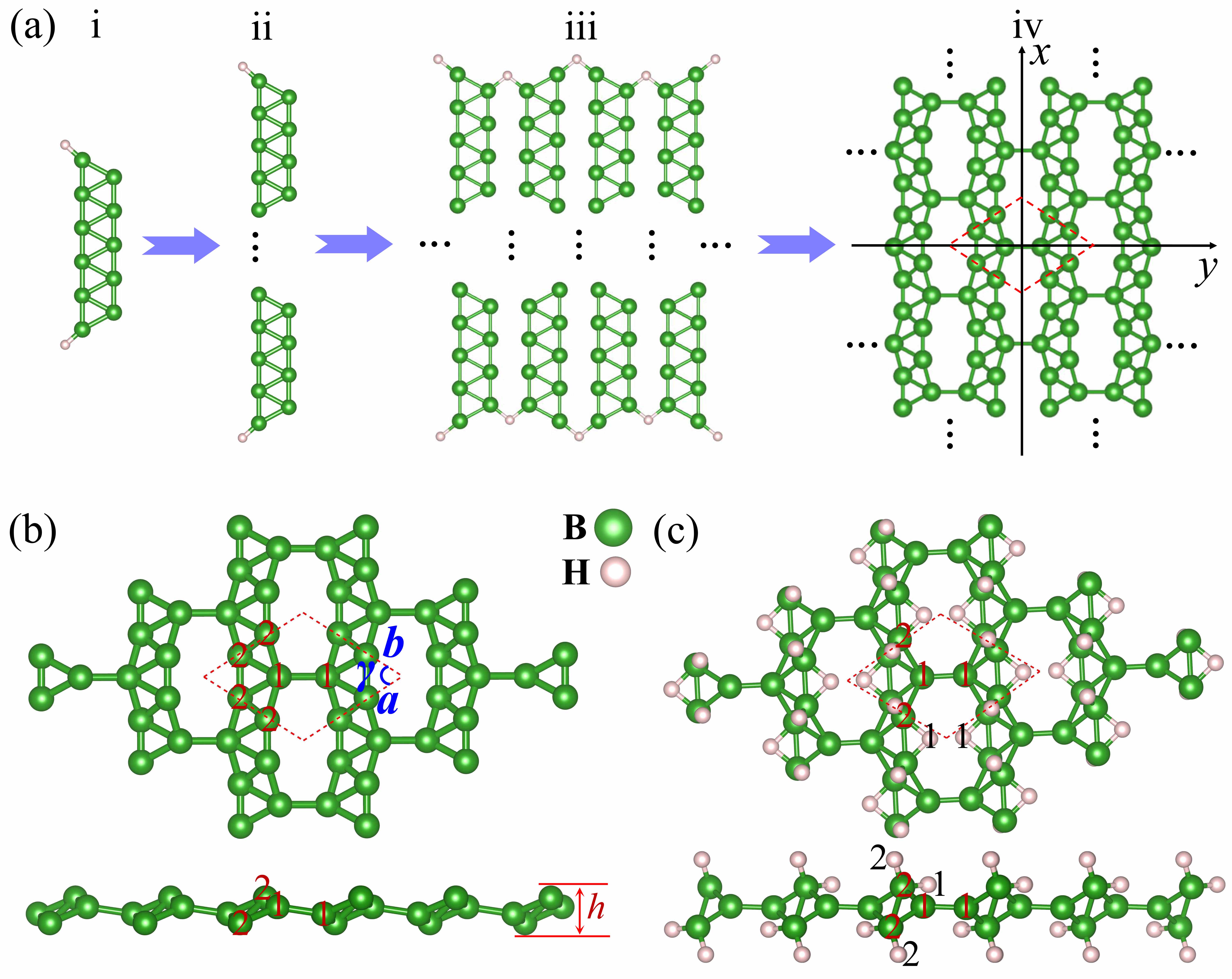}
\caption{(a) Schematic diagram for the self-assembly of boron cluster dihydrides: (i) isolated boron cluster dihydrides, (ii) a long molecular wire formed as the polymer of boron cluster dihydrides, (ii) periodical arrangement of the ladder-like molecular wires along $y$ direction, (iv) formation of a nanosheet structure of ladder polyborene extending along $x$ and $y$ directions. The minimal repetitive unit is marked by the red rhomboid. 
(b) Top and side views for the 2D ladder polyborene with \textbf{\emph{a}}=\textbf{\emph{b}}=4.362 {\AA}, ${\gamma}=66.75^{\circ}$, \emph{h}=0.980 {\AA}. (c) Top and side views for the 2D ladder polyborane with \textbf{\emph{a}}=\textbf{\emph{b}}=4.433 {\AA}, ${\gamma}=65.57^{\circ}$, \emph{h}=1.547 {\AA}.
The primitive cell is highlighted by the red dashed line. The number 1 and 2 refer to two kinds of inequitable boron atoms (B$_1$ and B$_2$) and hydrogen atoms (H$_1$ and H$_2$).
}\label{structure}
\end{figure}

Located in the IIIA group in the periodic table, boron ($2s^2p^1$) has a various valence electronic configuration, forming rich bonding characters from 0D to 3D boron-based polymorphisms\cite{szwacki2021historical,penev2012polymorphism}. Thereinto, the various boron clusters including the ladder-like B$_n$H$_2$-clusters\cite{li2012elongation} are viewed as the basic building block of 2D boron-based nanosheets, which are expected to possess potential applications in nanodevices\cite{alexandrova2006theoretical,li2017planar}.
As schematically exhibited in Fig.~\ref{structure}(a), starting from boron cluster dihydrides, through dehydrogenation they could polymerize into a long molecular wire, considered as polyene analogues. Then via further self-assembly of such ladder-like molecular wires, a 2D sheet of borophene that extends in both $x$ and $y$ directions is formed, where the minimal periodic unit extracted from the 2D sheet is identified and marked by the red rhomboid, termed as 2D ladder polyborene.

%%%%%%%%%%%%%%%%%%%%%%%%%
\begin{figure}
\includegraphics[width=4.5 in]{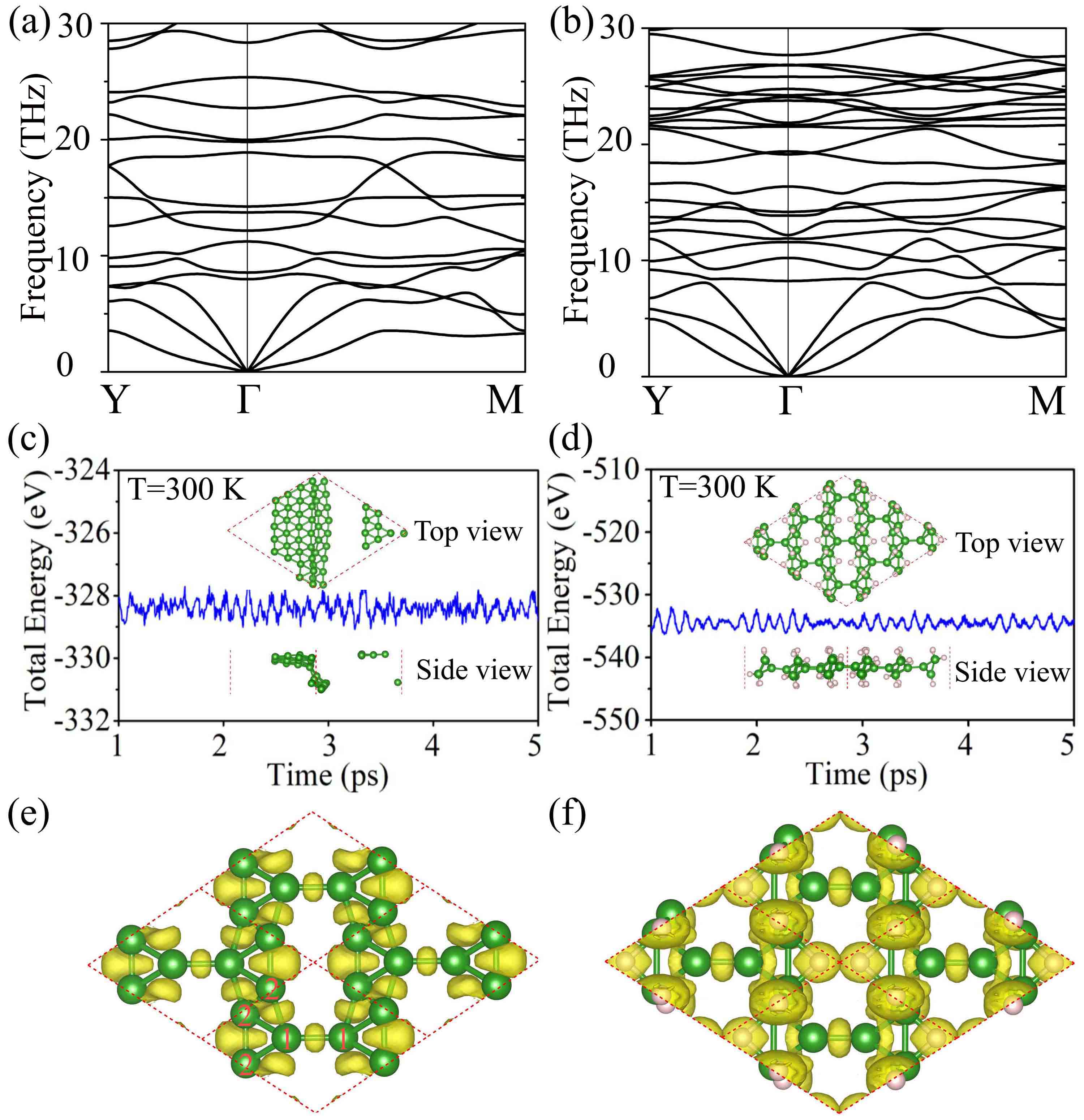}
\caption{(a), (b) The phonon band dispersions of 2D ladder polyborene and polyborane respectively. (c), (d) The total energies under AIMD simulation at 300 K with the snapshots of the crystal structures after 5 ps in the insets. (e), (f) Top views of ELFs with isosurface of 0.78 for ladder polyborene and polyborane.}\label{stability}
\end{figure}
%%%%%%%%%%%%%%%

The crystal structures of 2D ladder polyborene and its hydrogenated partner, 2D ladder polyborane, are displayed in Figs.~\ref{structure}(b) and (c), where each primitive cell contains six boron atoms that can be divided into two subgroups: B$_1$ and B$_2$.
Our calculations reveal that hydrogen atoms tend to absorb on the bridge site or top/bottom site of B$_2$ atom [see the details in section A of Supporting Information (SI)]. 
The calculated B-H bond lengths are 1.338 {\AA} for bridge-H and 1.193 {\AA} for top/bottom-H, in agreement with those of the borophane in experiment\cite{li2021synthesis}. These bond lengths are comparable with three-center-two-electron ($3c$-$2e$) B-H-B bond length in diborane\cite{laszlo2000diborane} and two-center-two-electron ($2c$-$2e$) B-H bond length in borane\cite{lipscomb1977boranes}, indicating distinctive covalent bonds in ladder polyborane. The formation energy of polyborane is about -0.587 eV per boron atom, much lower than that of rect-2H borophane\cite{pang2022first} [see details in section A of SI], implying the potential for experimental synthesis.
Besides, the Bader charge analysis shows that the boron atoms are positively charged while the hydrogen atoms are negatively charged indicating the hydride nature of ladder polyborane [see Fig. S2 in section B of SI].

According to the phonon spectra in Figs.~\ref{stability}(a) and (b), both 2D ladder polyborene and polyborane are dynamically stable under low temperature. To further estimate their thermostability, we performed ab initio molecular dynamics (AIMD)simulations under room temperature as shown in Figs.~\ref{stability}(c) and (d).
It is found that the structure of ladder polyborene breaks down into hexagon-packed pieces, which means that freestanding ladder polyborene is unstable at 300 K temperature. In contrast, the structure of ladder polyborane remains nearly unchanged under room temperature, and further calculation shows it can even maintain under 1000 K [see Fig. S3 in section C of SI].
This ultra-high thermostability of ladder polyborane mainly derives from various hydrogenation patterns and bonding nature.
As demonstrated in Figs.~\ref{stability}(e) and (f), the electron local functions (ELFs) for ladder polyborene reflects characteristic bond features of 2D boron sheet \cite{ou2021emergence} that involve the localized $2c$-$2e$ $\sigma$-bond between B$_1$ atoms and delocalized $mc$-$2e$ ($m>2$) bond among B$_2$ atoms. After hydrogenation, the ELF between two B$_1$ atoms obviously rises, resulting in a sturdier $\sigma$-bond. Meanwhile, the ELF among B$_2$ atoms becomes more extended along the $x$ direction, indicating a prominent multi-center bond.
All performances of electrons after hydrogenation contribute to promoting the thermodynamical stability of ladder polyborane as expected.
Besides, we also found that the oxidation of ladder polyborane is largely suppressed due to the protection of hydrogenation on both sides, thus it will demonstrate excellent thermostability under ambient conditions [see Fig. S4 in section D of SI].

\begin{figure}
\includegraphics[width=4.5 in]{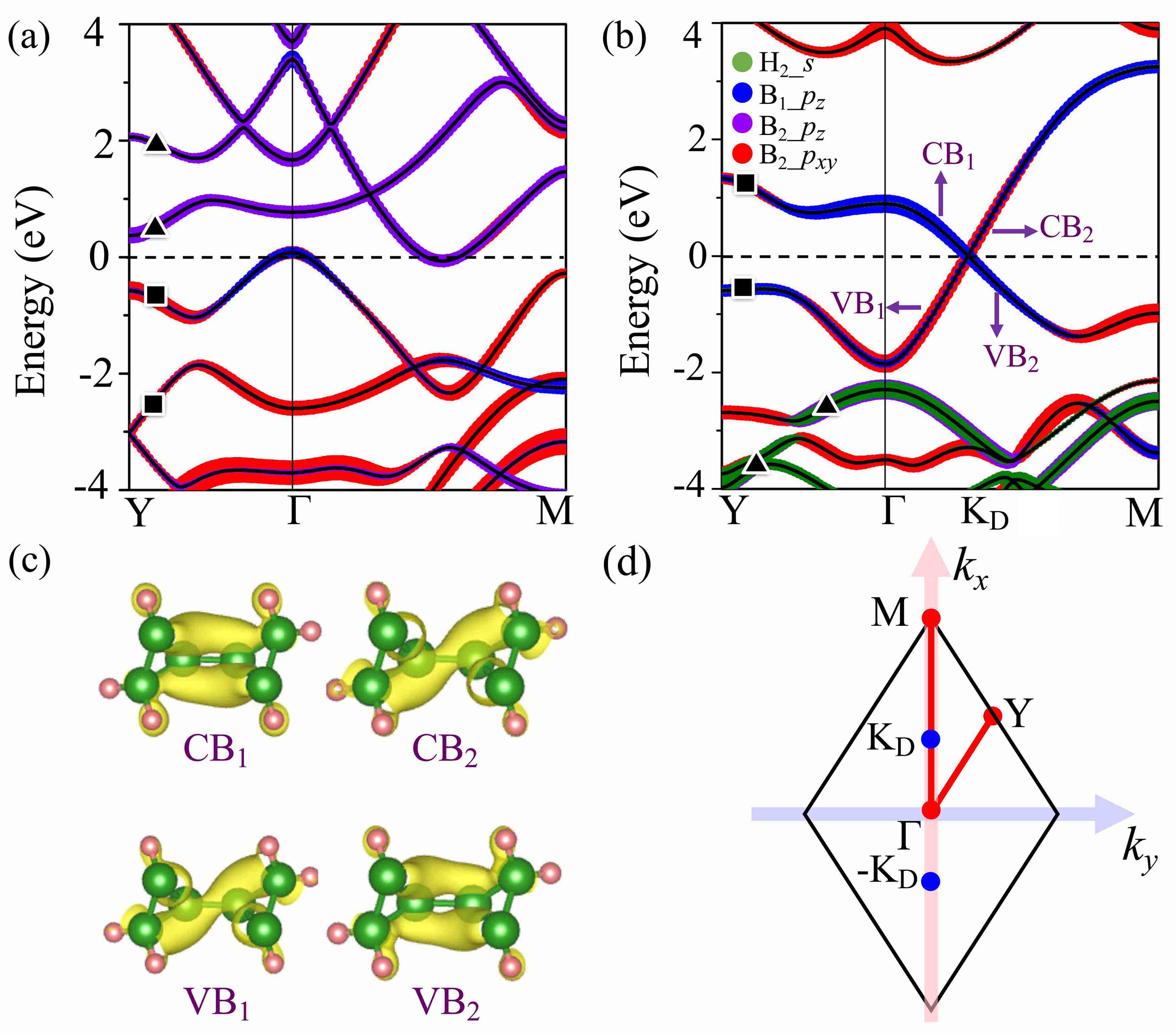}
\caption{(a), (b) The orbital-resolved band structures of ladder polyborene and polyborane, respectively.
(c) The partial charge density of the four patterns CB$_1$, CB$_2$, VB$_1$ and VB$_2$ marked in (b). (d) The Brillouin zone with high-symmetric lines and points. The K$_{\rm D}$ (0.152, -0.152) and -K$_{\rm D}$ indicate the locations of Dirac points.}\label{band}
\end{figure}

The Figs.~\ref{band} (a) and (b) demonstrate the orbital-resolved band structures of 2D ladder polyborene and polyborane, respectively.
Thereinto, the ladder polyborene presents conventional semimetal characteristics, whose valence band minimum (VBM) at the $\Gamma$ point overlaps with the conduction band maximum (CBM) along the $\Gamma$M path. 
Different from known borophene allotropes with metallic\cite{kim2015observation} or semiconducting nature\cite{jacsxu2017two,zhang2020fully}, the conventional semimetal band structure in ladder polyborene may lead to distinctive magnetic transport phenomena, such as unsaturated magnetic resistance.\cite{ali2014large}. 
In contrast, the ladder polyborane showcases an extremely clean scenario of Dirac semimetal, wherein the CBM and VBM linearly touch at the K$_{\rm D}$ point of the $\Gamma$M path, generating an anisotropic Dirac fermion at the Fermi level.
Therefore, we conclude that the hydrogenation not only largely enhances the thermodynamical stability of ladder polyborene, but also dramatically modifies the electronic structure by transforming it from normal semimetal into topological Dirac semimetal.

%%%%%%%%%%%%%%%%%%
%%%%%%%%%%%%%%%%%%%%%%%

The underlying mechanism can be further uncovered by orbital projections.
It is found that the lower conduction bands of ladder polyborene [labeled by black triangles in Fig.~\ref{band}(a)] are comprised of B$_2$'s $p_z$ and $p_{xy}$ orbitals while its higher valence band [represented by black squares in Fig.~\ref{band}(a)] are composed of a mixture of B$_1$'s $p_z$ and B$_2$'s $p_{xy}$ orbitals.
Contrastively, the lowest conduction and highest valence bands of ladder polyborane are mainly dominated by a mixture of B$_1$-$p_z$ and B$_2$-$p_{xy}$ orbitals.
Since the H$_2$ atom absorbs on the top site/bottom sites of B$_2$ atom, H$_2$'s $s$ orbital will strongly bond with B$_2$' $p_z$ orbital.
Resultantly, the two lowest conduction bands of ladder polyborene will be filled up by electrons, and thus become deep valence bands below -2 eV in ladder polyborane, as highlighted by the black triangles in Fig.~\ref{band}(b).
Meanwhile, the two highest valence bands of ladder polyborene are pushed up, leading to the emergence of the Dirac cone at the Fermi level in ladder polyborane, as highlighted by the black squares in Fig.~\ref{band}(b).
Moreover, the orbital ingredient and bonding nature around the Dirac cone can also be testified by the partial charge density in Fig.~\ref{band}(c).
One can observe that patterns of CB$_1$ and VB$_2$ share similar charge densities, which mainly accumulate at the center of B$_1$-B$_1$ bond, showing typical $\pi$ bonding-state of \emph{p}$_z$ orbital, whereas patterns of CB$_2$ and VB$_1$ present $\sigma$ bonding-state of $p_{x,y}$ orbital, in accord with above orbital projection analysis.

To further reveal the anisotropy and tilting effect of Dirac fermion in ladder polyborane, a low-energy effective $k{\cdot}p$ model around K$_{\rm D}$ and K$_{\rm D}^{'}$(-K$_{\rm D}$) points is developed [see section E of SI for details], and written as,
\begin{eqnarray}
H={v}_{y}{\sigma }_{y}{k}_{y}+\tau_{z}{v}_{x}{\sigma}_{z}{k}_{x}+\tau_{z}{w}_{x}{k}_{x}{\sigma }_{0},\label{Hami}
\end{eqnarray}
where $\sigma_0$ is the identity matrix, $\sigma_{x,y,z}$ are Pauli matrices and $\tau_{z}$ refers to valley index (K$_{\rm D}$/K$_{\rm D}^{'}$) shown in Fig.~\ref{band}(d).
The parameters ${v}_{x}=7.35\times 10^5$ $m/s$, ${v}_{y}=3.97\times 10^5$ $m/s$ and ${w}_{x}=1.91\times 10^5$ $m/s$ are obtained by fitting $k\cdot p$ bands with density functional theory results.

\begin{figure*}
\includegraphics[width=6.5 in]{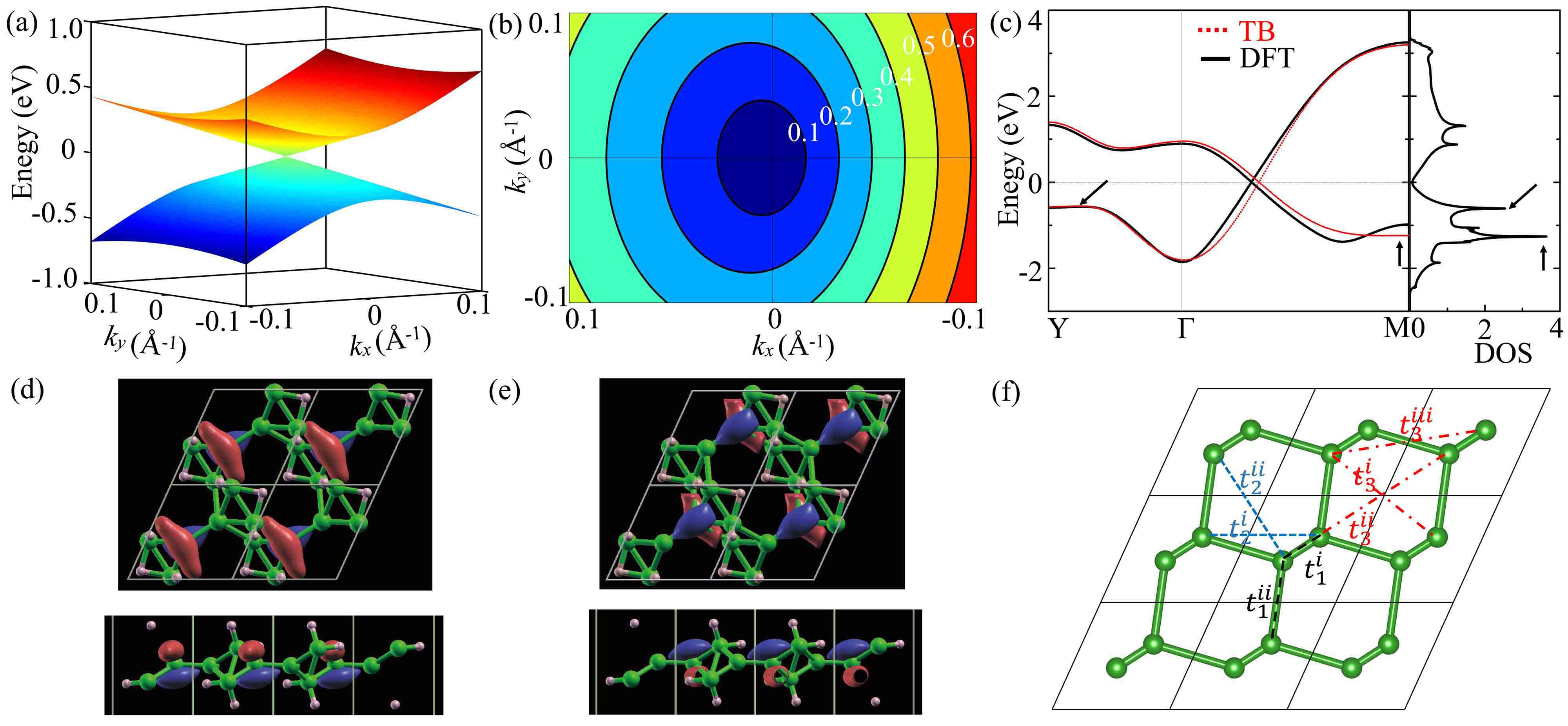}
\caption{The Dirac cone around K$_{\rm D}$ point in (a), with its constant energy contour displayed in (b). (c) The band structures from TB model (red line) and DFT calculation (black line) with DOS. The black arrows point to the van Hove singular points. (d) and (e) are distributions of two localized Wannier orbitals in our TB model. (f) The distorted hexagon lattice model consisting of B$_1$ atoms with WFs is displayed, including two types of first-order neighbor hoppings ($t_1^{i,ii}$), two types of second-order neighbor hoppings ($t_2^{i,ii}$), and three kinds of third-order neighbor hoppings ($t_3^{i,ii,iii}$).}\label{tbdft}
\end{figure*}

From Eq. (\ref{Hami}), one can readily obtain the dispersion relation as $E_{\eta}(k)={\eta}w_xk_x \pm (v_x^2k_x^2+v_y^2k_y^2)^{1/2}$ (${\eta}$=$\pm$1 stands for valley index), which describes an anisotropic Dirac cone with a tilting effect as clearly illustrated in Fig.~\ref{tbdft}(a). Specifically, the first two terms in Eq. (\ref{Hami}) dominate the topological properties of the Dirac cone and the ratio of parameters $r_{a}$=$v_y$/$v_x$ determines the anisotropy.
Unlike graphene ($r_{a}$=1) which hosts an isotropous Dirac cone, the $r_{a}$=0.54 in ladder polyborane indicates significant anisotropic dispersion as reflected by elliptical constant energy contour in Fig.~\ref{tbdft} (b).
The third term is the tilting term along the $k_x$ direction that breaks Lorentz invariance of Dirac fermion. The degree of inclination is measured by $r_t$=$w_x/v_x$. The $r_t$=0.26 for ladder polyborane leads to visibly asymmetric dispersion between $+k_x$ and $-k_x$ direction, therefore the constant energy contour become an off-centered ellipse.
These features in band dispersion directly manifest in its transport properties\cite{tan2021anisotropic,TmPhysRevB.91.115135}, e.g. direction-dependent Fermi velocity ($v_F$), which is 9.26 along $+k_x$, 5.44 along $-k_x$, and $3.97$ along $\pm k_y$ with a unit of $10^{5}$ $m/s$.
In addition, this anisotropic Dirac cone in ladder polyborane is further confirmed by a more accurate HSE06 functional [see Fig. S5 in section F of SI].

%%%%%%%%%%%%%%%%%

To gain more insights into the physics of ladder polyborane, we further construct a minimal tight-binding (TB) model to describe its band structure.
Since the lowest conduction and highest valence bands are well-separated from the residual parts, it allows us to utilize two localized Wannier functions (WFs) to build such a TB model.
In Figs.~\ref{tbdft} (d) and (e), the WFs localized on the B$_1$ atoms are obtained, which demonstrates features of $p_z$-orbital but with obvious anisotropy.
After considering the on-site energy (${\epsilon}_i$), the first, second, and third-order neighbor hopping terms ($t_{1,2,3}$), the Hamiltonian is written as,
\begin{equation}\label{tb1}
\begin{split}
H&=\sum_{i} {\epsilon}_i c_{i}^{\dag}c_{i}+\sum_{<ij>} t_{1}c_{i}^{\dag}c_{j}\\
& +\sum_{<<ij>>} t_{2}c_{i}^{\dag}c_{j}+\sum_{<<<ij>>>} t_{3}c_{i}^{\dag}c_{j}.
\end{split}
\end{equation}

In Fig.~\ref{tbdft} (f), a lower spatial symmetry of distorted honeycomb lattice leads to distinctive hopping parameters. Based on these fitting parameters in Table ~\ref{table1}, we successfully reproduce the conduction and valence bands of ladder polyborane within the whole Brillouin zone. As comparably illustrated in Fig.~\ref{tbdft} (c), in addition to the Dirac cone at the Fermi level, there exist multiple van Hove singular points near the Fermi level which contribute diverging density of states (DOS) as highlighted by black arrows.
The first DOS peak locates only 0.6 eV below the Fermi level, corresponding to 1.7$\times$10$^{14}$ cm$^{-2}$ hole doping, which is experimentally reachable via gate voltage, and thus can be a platform for exploring many-body phenomena as recently observed in deeply-doped graphene\cite{rosenzweig2020overdoping}.

It is noteworthy that despite many 2D Dirac materials being theoretically predicted\cite{xu2016hydrogenated,zhang2016dirac,zhang2019borophosphene}, such pure Dirac electronic state that can be perfectly described by a minimal two-band TB model is very rare. Therefore, the ladder polyborane may provide a concise and flexible material platform for future exploration of anisotropic and tilted Dirac electrons. Here, in analogous to graphene, we also explored the electronic structure of zigzag nanoribbon of ladder polyborane. The flat edge states with distinctive magnetic configurations are discovered [see Fig. S6 of section G in SI]

	\begin{table}
		\caption{Fitting parameters of TB model in the unit of eV.}
		\label{table1}
		\setlength{\tabcolsep}{3.6pt}
		\renewcommand{\arraystretch}{1.5}
		\begin{tabular}{cccccccc}
    \hline
    \hline
  $\epsilon_i$ & $t_1^i$ & $t_1^{ii}$ & $t_2^{i}$ & $t_2^{ii}$ & $t_3^{i}$ & $t_3^{ii}$ & $t_3^{iii}$ \\
    \hline
  -3.040 & -0.701 & 0.886 & -0.176 & -0.034 & 0.184& -0.085 & -0.014 \\
  \hline
    \hline
  \end{tabular}
	\end{table}

On the other hand, the manipulation of Dirac semimetal states and exploring their rich physical properties are of great significance. It's known that the introduction of spin-orbit coupling effect, electric field, and CPL-field into 2D massless Dirac fermion will open up a nontrivial band gap, and give birth to diverse topological quantum states\cite{ygyao2014PhysRevLett.112.106802}. In fact, despite several proposed 2D Dirac materials, the manipulation of Dirac fermions by external fields still remains a great challenge\cite{liu2009bandgap,sahu2017band}.
Since Dirac point in ladder polyborane is protected by the coexistence of spatial inversion ($\mathcal{P}$) and time-reversal ($\mathcal{T}$) symmetries, as revealed by the nonzero Z$_2$ number,\cite{fan2018cat} breaking either $\mathcal{P}$ or $\mathcal{T}$ naturally generates a band gap and realizes massive Dirac fermions. A convenient way is applying vertical electric field breaking $\mathcal{P}$ or circularly polarized light (CPL) breaking $\mathcal{T}$.
For ladder polyborane, due to its buckled boron atoms, it is possible to realize band-gap engineering and topological phase transitions directly by imposing external fields as exhibited in the following.

\begin{figure}
\includegraphics[width=4.5 in]{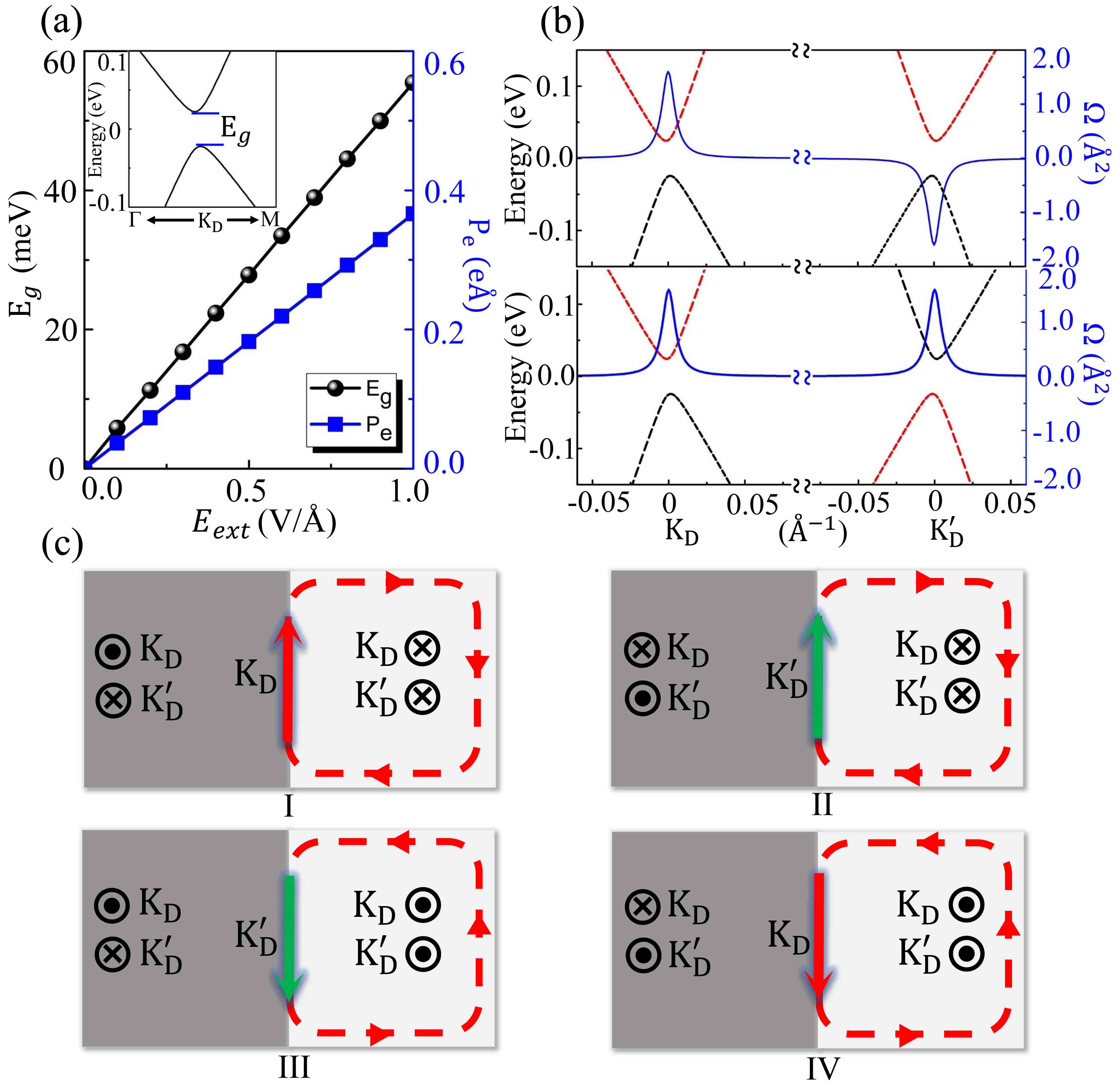}
\caption{(a) The evolution of band gap (E$_g$) and electric polarization (P$_e$) with increasing vertical electric field. Insets are the band structure under the external vertical electric field of 0.8 V/{\AA}. (b) Energy bands and Berry curvature around K$_{\rm D}$ and K$_{\rm D}^{'}$ under broken $\mathcal{P}$-symmetry (top panel) and under broken $\mathcal{T}$-symmetry (bottom panel). (c) The schematic illustrations of the TDWs driven by multi fields. The directions of the arrows perpendicular to the page represent the sign of the valley-specific Hall conductivities. The circular dashed line indicates the chiral edge state. The red and green solid arrows indicate the topological interfacial states with chirality and valley index.
}\label{efiled}
\end{figure}

Firstly, we discuss the effect of a vertical electric field on ladder polyborane. In Fig.~\ref{efiled} (a), an available electric field can induce a considerable band gap at K$_{\rm D}$ point, generating a massive Dirac fermion.
The band gap, $E_{g}$, depends linearly on the electric field strength ($E_{ext}$) with a ratio of 0.054 e{\AA}.
Meanwhile, an electronic polarization ($P_e$) is unavoidably induced by the external electric field, which in turn partially shields this field. In the inset of Fig.~\ref{efiled}(a), a band gap of 45 meV is obtained when $E_{ext}$=0.8 V/{\AA}, which may have promising applications in low-temperature field effect transistor (FET)\cite{drummond2012electrically,tao2015silicene}.

This significant impact of the electric field with nontrivial topological properties can be interpreted by symmetry analysis on the $k{\cdot}p$ model.
Since the vertical electric field induces distinctive effective potential on different atomic orbitals, the $\mathcal{P}$-symmetry breaks, and the Semenoff mass term $m_0{\sigma }_{x}$ emerges\cite{PhysRevLett.53.2449}. Therefore, the $k{\cdot}p$ model is written as,
\begin{eqnarray}
H={v}_{y}{\sigma }_{y}{k}_{y}+\tau_{z}{v}_{x}{\sigma}_{z}{k}_{x}+\tau_{z}{w}_{x}{k}_{x}{\sigma }_{0}+m_0{\tau}_{0}{\sigma }_{x}.
\end{eqnarray}
It portrays a massive Dirac fermion at K$_{\rm D}$/K$_{\rm D}^{'}$ with the $m_0$=0.027$E_{ext}$ according to our first principles calculation.
Obviously, the sign of mass could be modified by flipping the electric field direction.
It's known that the broken $\mathcal{P}$-symmetry allows the emergence of large Berry curvature distribution centered around K$_{\rm D}$/-K$_{\rm D}$.\cite{xiao2007valley} The Berry curvature, $\boldsymbol{\Omega} (\textbf{\emph{k}})$, is defined by
$\boldsymbol{\Omega} (\textbf{\emph{k}})=\nabla_\textbf{\emph{k}}\times \bra{u(\textbf{\emph{k}})} i\nabla_\textbf{\emph{k}} \ket{u(\textbf{\emph{k}})} $, where $\ket{u(\textbf{\emph{k}})}$ is the periodic part of the Bloch function.
Accordingly, the valley-specific Hall conductivity ($\sigma_H^{\eta}$, ${\eta=\pm 1}$ stands
for different valleys), Hall conductivity ($\sigma_H$), and valley Hall conductivity ($\sigma_H^V$) are defined as,\cite{PhysRevB.84.075418}
\begin{equation}\label{Hall}
\begin{split}
\sigma_H^{\eta}=\int \frac{dk^2} {{2\pi}^2}\boldsymbol{\Omega} (\textbf{\emph{k}}) \\
\sigma_H =  \sigma_H^{K_D}+ \sigma_H^{K_D^{'}}  \\
\sigma_H^V = \sigma_H^{K_D}-\sigma_H^{K_D^{'}}.
\end{split}
\end{equation}
%%%%%%%%%%%%%%%%%%%%%%%%%%%%

Interconnected by $\mathcal{T}$-symmetry, two valleys host the same band gap but opposite Berry curvature as displayed on the top panel of Fig.~\ref{efiled}(b), which leads to $\sigma_H^{\eta}$=$\frac{1}{2} \eta \textit{sgn} (E_{ext})$ in the unit of $\frac{e^2}{h}$, where $sgn$ refers to the sign function.
Thus, the carriers in different valleys will flow to opposite transverse edges when an in-plane electric field is applied, namely the valley Hall effect (VHE).\cite{yamamoto2015valley,PhysRevLett.108.196802}
Correspondingly, we can readily obtain $\sigma_H$=0, and $\sigma_H^{V}$=$\eta$\textit{sgn}($E_{ext}$).
The value of $\sigma_H^{V}$ is quantized, namely the quantum valley Hall effect (QVHE), \cite{DJPhysRevB.84.195444,EzawaPhysRevB.87.155415,PhysRevX.9.031021} and its sign can be flexibly changed by flipping the electric field direction.

Secondly, shining light radiation by a circularly polarized light (CPL) can break $\mathcal{T}$-symmetry and lead to exotic topological states.\cite{Kitaprb2011,li2018floquet,he2019floquet}
Based on the Floquet theory, the CPL with frequency $\omega$ and amplitude $A_0$ is expressed as $\boldsymbol{A}\left(t\right)=A_{0}\left(cos(\omega t), \xi sin(\omega t)\right)$. The $\xi =1 (-1)$ denotes the right-handed (left-handed) CPL.
Considering the $\boldsymbol{A}(t)$ coupling with  Hamiltonian in Eq. (\ref{Hami}), we finally deduce an effective Floquet Hamiltonian as,
\begin{equation}\label{Floquet-eff}
\begin{split}
H_{F}\left(\boldsymbol{k}\right)&=\tau_{z}v_{x}k_{x}\sigma_{z}+v_{y}k_{y}\sigma_{y}\\\
&+\tau_{z}\omega_{x}k_{x}\sigma_{0}+\lambda_{k}\xi\tau_{z}\sigma_{x},
\end{split}
\end{equation}
where $\lambda_{k}=\frac{v_{x}v_{y}\left(eA_{0}\right)^{2}}{\hbar\omega}$ [see details in section H of SI].
The last term $\lambda_{k}\xi\tau_{z}\sigma_{x}$ is equivalent to famous Haldane mass term\cite{haldaneRevModPhys.89.040502}. It generates an opposite band gap but the same Berry curvature for two valleys as displayed on the bottom panel of Fig. ~\ref{efiled}(b), which contributes to $\sigma_H^{\eta}$=$\frac{1}{2} \xi$. Resultantly, $\sigma_H$=$\xi$ and $\sigma_H^V$=0, a quantized Hall conductivity is readily obtained.
This nontrivial state motivated by CPL is called the Floquet Chern insulator (FCI), where the sign of $\sigma_H$ can be flipped by choosing different chirality of CPL.
Specifically, the energy of CPL is usually chosen as the band width\cite{saha2016PhysRevB.94.081103}, which is about 7.8$\times$10$^{15}$ Hz for ladder polyborane, then a band gap of 50 meV is able to be opened under typical experimental light field strength of $eA_0$=0.07 ${\AA}{^{-1}}$ as displayed in Fig. ~\ref{efiled}(b).

%%%%%%%%%%%
So far we have exhibited ladder polyborane as an ideal Dirac semimetal with multi-field-tunable band gap to raise various non-trivial topological insulating states. A large superiority of QVHE and FCI driven by external fields is that their topological indexes could be easily regulated by flipping field directions. Based on this fact, we proposed the concept of a topological domain wall (TDW) \cite{PhysRevB.89.085429,maoyuanPhysRevB.93.155412}, where a uniformed electric field is applied on the left side and a CPL field is exerted on the right side. Since two sides host different topological charges ($\sigma_H^{\eta}$), it's natural to define the topological invariant for such TDW as,
\begin{equation}\label{DW}
\begin{split}
\delta \sigma_H^{\eta}= \sigma_{H,L}^{\eta}-\sigma_{H,R}^{\eta} \\
\end{split}
\end{equation}
where the subscript $L(R)$ stands for the left (right) side of TDW and $\eta$=K$_{\rm D}$/K$_{\rm D}^{'}$.
Similarly, the $\delta \sigma_H$ and $\delta \sigma_H^V$ are defined as the sum and difference of ${\delta}\sigma_H^{K_D}$ and ${\delta}\sigma_H^{K_D^{'}}$, respectively. The $\delta \sigma_H$ identifies the number of net chiral conducting states while the $\delta \sigma_H^V$ stands for the number of net valley-polarized conducting states on the interfacial region.
Considering all possible scenarios, we obtained four kinds of TDWs as schematically displayed in Fig. \ref{efiled} (c).
For phase-I, when the positive electric field and the left-handed CPL-filed are exerted, one can find
$\delta \sigma_H$=$\delta \sigma_H^V$=${\delta}\sigma_H^{K_D}$=$1$. These non-zero topological numbers refer to a right-chiral and meanwhile K$_{\rm D}$-valley polarized interfacial state. More significantly, by simply reversing the electric field direction on phase-I, we can obtain phase-II with $\delta \sigma_H$=$-\delta \sigma_H^V$=${\delta}\sigma_H^{K_D^{'}}$=$1$, which supports a right-chiral but K${_D^{'}}$-valley polarized interfacial state, namely realizing the manipulation of the degree of valleys. Moreover, by choosing different CPL-field, the chirality of interfacial state can be modified, as exhibited in phase-III and phase-IV, respectively.
Therefore, via multi-filed driven TDW structure in ladder polyborane, it is flexible to generate topological interfacial state with regulable chirality and valley index, forming the basis for valley-based electronics applications.

\section{CONCLUSIONS}
In summary, we propose 2D ladder polyborane as an ideal Dirac semimetal that shares high thermal stability and tunable electronic properties. A low-energy effective $k{\cdot} p$ model and a minimal TB model are developed to depict the anisotropic and tilted Dirac cone as well as van Hove singular points. Moreover, we reveal that ladder polyborane offers a versatile material platform to realize abundant TDW states driven by electric field and CPL field. It is believed this kind of light-element material, ladder polyborane, with flexibly tunable Dirac fermions can provide an excellent avenue to study the fascinating properties of 2D Dirac material in future applications.

\section{METHODS}
Our first-principles calculations are performed based on density functional theory (DFT) as implemented in the Vienna $ab$-$initio$ simulation package (VASP)\cite{vasp1,vasp2}. The exchange-correlation part was described with the generalized gradient approximation (GGA) in the form proposed by Perdew-Burke-Ernzerhof (PBE) functional\cite{gga1}.
The 450 eV cutoff energy was exploited and the Brillouin zone was sampled with a 16$\times$16$\times$1 $\Gamma$-centered Monkhorst-Pack k-points grid.
The energy precision was set to 10$^{-5}$ eV per atom, and the atomic positions were fully relaxed until the maximum force on each atom was less than 10$^{-2}$ eV/\AA. The vacuum space of 20 {\AA} is used to avoid interactions between adjacent layers. Moreover, to assess the structural stability of ladder polyborene and polyborane, the phonon dispersions are calculated by using DFPT method \cite{DFPT1}. The structural stability is confirmed by ab initio molecular dynamics (AIMD) and phonon spectra using a 3$\times$3$\times$1 supercell. The Wannier orbitals are obtained by the Wannier90 code\cite{wannier1}.

\section{Supporting Information}
The Supporting Information includes: (A) The hydrogenation patterns of ladder polyborene, (B) Bader charge analysis of ladder polyborane, (C) Thermodynamic stability of ladder polyborene and polyborane, (D) Adsorption of O$_2$ molecular on ladder polyborane, (E) Low energy effective $k{\cdot}p$ model of ladder polyborane, (F) HSE06 correction of the band structure of ladder polyborane, (G) Electronic structures of zigzag nanoribbon of ladder polyborane, (H) Photoinduced Floquet Chern insulator in ladder polyborane.

\section{Notes}
The authors declare no competing financial interest.

\begin{acknowledgement}
This work was supported by the National Natural Science Foundation of China (Grant Nos. 11922401, 12204330), the National Key R$\&$D Program of China (Grant No. 2020YFA0308800), the China Postdoctoral Science Foundation (Grant Nos. 2020M680011, 2021T140057), China National Postdoctoral Program for Innovative Talent (Grant No. BX20220367), and the Sichuan Normal University for financial support (No. 341829001). The numerical computations were performed on Hefei advanced computing center, and this research was also supported by the High Performance Computing Center of Sichuan Normal University, China. 
\end{acknowledgement}

%%%%%%%%%%%%%%%%%%%%%%%%%%
%%%%%% reference%%%%%%%%%%

\providecommand*\mcitethebibliography{\thebibliography}
\csname @ifundefined\endcsname{endmcitethebibliography}
  {\let\endmcitethebibliography\endthebibliography}{}

%%%%%%%%%%%%%%%%
%%%%%%%%%%%%%%%%
%\bibliography{refs}

\end{document}